\documentstyle{article}
\include{epsf}
\include{psfig}
\makeatletter
\def\gsim{\compoundrel>\over\sim}

\def\compoundrel#1\over#2{\mathpalette\compoundreL{{#1}\over{#2}}}
\def\compoundreL#1#2{\compoundREL#1#2}
\def\compoundREL#1#2\over#3{\mathrel
         {\vcenter{\hbox{$\m@th\buildrel{#1#2}\over{#1#3}$}}}}
\makeatother
\newcommand{\dslash}{\! \not \! \partial}
\makeatletter
\leftmargin\leftmargini
\newdimen\Parindent\newdimen\Parskip
\Parindent=\parindent\Parskip=\parskip
\def\@oddhead{}\def\@evenhead{}
\def\@oddfoot{\rm\rightmark \hfil Page \thepage}
\def\@evenfoot{\rm\leftmark Page \thepage \hfil}

\newdimen\Parindent\newdimen\Parskip
{\Parindent=\parindent\Parskip=\parskip
\parindent0pt\parskip3mm\columnsep11mm}
{\parindent\Parindent\parskip\Parskip}
{\Parindent=\parindent\Parskip=\parskip
\parindent0pt\parskip3mm\columnsep11mm}
{\parindent\Parindent\parskip\Parskip}

\gdef\abstract#1{\gdef\@abstract{#1}}

\def\maketitle{\par
 \begingroup
 \setcounter{footnote}{0}\setcounter{page}{1}
 \let\save@thefootnote=\thefootnote
 \let\save@makefnmark=\@makefnmark
 \def\thefootnote{\fnsymbol{footnote}}
 \def\@makefnmark{\hbox
 to 2mm{$\m@th^{\@thefnmark}$\hss}}
\kern1mm 
\@maketitle
 \@thanks{}%
 \endgroup
 \let\maketitle\relax
 \let\@maketitle\relax
 \gdef\@thanks{}\gdef\@author{}\gdef\@title{}\let\thanks\relax}

\def\endtitlepage{
\@thanks{}%
\let\thefootnote=\save@thefootnote
\let\@makefnmark=\save@makefnmark
\setcounter{footnote}{0}\let\maketitle\relax\vskip 3mm\upshape
\gdef\@author{}\gdef\@title{}\global\let\thanks\relax
\global\let\@thanks\relax}

\def\@maketitle{\vbox to 40mm{\hsize\textwidth
 \linewidth\hsize \vfil \centering
 {\vskip 1em  \Large\bf \@title \par} \vskip 2em
{\begin{center} \large\bf \@author\end{center}\par}\vfil}
\hsize\textwidth \linewidth\hsize
\vskip 1em
\begin{center} ABSTRACT \end{center} \par
\begin{center}\parbox{140mm}{\@abstract}\end{center} \vskip 3em }

\renewcommand{\section}{\@startsection
{section}
{1}
{\z@}
{-1.0\baselineskip}
{0.1\baselineskip}
{\large\bf}}%
\renewcommand{\subsection}{\@startsection
{subsection}
{2}
{\z@}
{-1.0\baselineskip}
{0.1\baselineskip}
{\large\bf}}%
\renewcommand{\subsubsection}{\@startsection
{subsubsection}
{3}
{\z@}
{-1.0\baselineskip}
{0.1\baselineskip}
{\normalsize\bf}}%
\renewcommand{\paragraph}{\@startsection
{paragraph}
{4}
{\z@}
{-1.0\baselineskip}
{0.1\baselineskip}
{\normalsize\bf}}%
\renewcommand{\subparagraph}{\@startsection
{subparagraph}
{4}
{\z@}
{-1.0\baselineskip}
{0.1\baselineskip}
{\normalsize\bf}}%
\makeatother
\begin{document}
\title{The Top Quark and other Fermion Masses }
\author{Colin Froggatt \\
        {\em Glasgow University, Glasgow G12 8QQ, Scotland} \\
\vspace{0.3cm}
{\rm To be published
in the Proceedings of the Fifth Hellenic
School and Workshops on Elementary Particle Physics, Corfu,
3 - 24 September 1995.}
       }
\vspace{-0.5cm}
\abstract{Recent developments on approaches to the quark lepton mass
problem are reviewed. In particular we discuss dynamical calculations
of the top quark mass at (a) the infrared quasifixed point of
the Minimal Supersymmetric Standard Model renormalisation group
equations, and (b) a strongly first order critical point of the
Standard Model effective potential. The phenomenology of symmetric
mass matrix ans\"{a}tze with texture zeros at the unification scale
is also considered. The underlying chiral symmetry presumed
responsible for the fermion mass hierarchy is discussed, with particular
reference to superstring based models.}

\maketitle
\section{Introduction}\label{sec:intro}
The pattern of observed quark and lepton masses, their
mixing and three generation structure form the major
outstanding problem of particle physics. The fermion masses
and mixing angles derive from Yukawa couplings, which are
arbitrary parameters within the Standard Model (SM).
Their experimental values provide our best clue to the
physics of flavour beyond the SM. The main features
requiring explanation are the following.
\begin{enumerate}
\item The large mass ratios between generations:
$m_u \ll m_c \ll m_t ;
\quad
m_d \ll m_s \ll m_b ;
\quad
m_e \ll m_{\mu} \ll m_{\tau}$.
\item The large mass splitting within the third (heaviest)
generation:
$m_{\tau} \sim m_b \ll m_t$.
\item The smallness of the off-diagonal elements of
the quark weak coupling matrix $V_{CKM}$.
\end{enumerate}
The above features constitute the charged fermion mass
and mixing hierarchy problem. The masses range over
five orders of magnitude, from $\frac{1}{2}$ MeV for the
electron to $\sim$ 200 GeV for the top quark. It is only
the top quark which has a mass of order the electroweak scale
$\langle\phi_{WS}\rangle$ = 246 GeV and a Yukawa coupling
of order unity. All of the other fermion masses are
suppressed relative to the natural scale of the SM.
The main problem in understanding the fermion spectrum
is not why the top quark is so heavy but rather
why the electron is so light. Indeed, as the top quark mass
is the dominant term in the fermion mass matrix, it is likely
that its value will be understood dynamically before those
of the other fermions.

As is well known, each generation
\mbox{(u,d,e,$\nu_e$)}, \mbox{(c,s,$\mu,\nu_\mu$)} and
\mbox{(t,b,$\tau,\nu_\tau$)}
forms an anomaly free representation of the SM gauge
group (SMG). The LEP measurements of the invisible partial
decay width of the Z boson show that there are just three neutrinos
of the usual type with masses less than $M_{Z}/2$
or more precisely \cite{pdg}
\begin{equation}
N_\nu = 2.985 \pm 0.023 \pm 0.004
\end{equation}
This result is naturally interpreted to imply that there
are just three generations of quarks and leptons,
since neutrinos are purely left-handed and massless in the SM.
It is of course possible to consider a fourth generation
with a heavy neutrino at the electroweak scale \cite{hillpas}
(e.~g.~by introducing one isosinglet right handed neutrino).
It is also possible to add a fourth generation of quarks
without accompanying leptons; in this case it is necessary to
cancel the associated contributions to the gauge anomalies
against those of some further particles. In fact the
uncancelled quark anomalies make up a natural unit that
is rather easily cancelled by a kind of techniquark.
This can be done by adding an SU(5) component to the SMG
togther with a generation of SU(5) "quarks". The resulting
model \cite{douglas} has many features similar to the SM,
including a natural generalisation of the SM charge
quantisation rule. We do not consider these unconventional
possibilities further here and just assume the usual picture
of three SM generations.

As mentioned above there are no right-handed weak
isosinglet neutrino states $\nu_R$ in the SM;
so the Higgs mechanism cannot generate a
neutrino mass term. However, in extensions of the SM,
it is straightforward to generate Majorana
mass terms connecting the left-handed weak
isodoublet neutrinos of the SM with
the corresponding set of right-handed
weak isodoublet anti-neutrinos. These
Majorana mass terms break weak isospin
by one unit ($\Delta t = 1$) as well
as lepton flavour conservation.
Such a $\Delta t = 1$ mass term can be
generated by:
(i) the exchange of the the usual Higgs tadpole
$\langle\phi_{WS}\rangle$ twice, via a superheavy
lepton $L^0$ intermediate state having the same
gauge quantum numbers as $\nu_R$ (i.~e.\ neutral)
under the SM \cite{fn2,seesaw}; or
(ii) the exchange of a single weak isotriplet
Higgs tadpole \cite{gelmini}.
Method (i) has become known as the see-saw
mechanism, since it generates a neutrino mass scale of
\mbox{$\langle\phi_{WS}\rangle^2/M_{L^0}$},
suppressed by a factor of
\mbox{$\langle\phi_{WS}\rangle/M_{L^0}$}
relative to the natural charged
fermion mass scale of
\mbox{$\langle\phi_{WS}\rangle$.}
Some recent applications of the see-saw mechanism
to make model predictions of neutrino masses and
mixings will be found in other contributions to this meeting.
Here we shall concentrate on the charged fermion mass problem.

Firstly, in section \ref{sec:top}, we report on recent
attempts to determine the top quark mass $m_t$
(or more generally third generation masses) dynamically
within the SM or the Minimal Supersymmetric Standard
Model (MSSM). We then review some different
approaches to the fermion mass and mixing hierarchy problem.
In section \ref{sec:ansatz} we consider ans\"{a}tze for the
fermion mass matrices. By imposing symmetries and
texture zeros on the matrices, it is possible to obtain
testable relations between the masses and mixing angles.
(Texture zeros are small mass matrix elements which can
be neglected to leading order.) In this approach the
mass hierarchy is usually simply imposed, by fitting
parameters in the ans\"{a}tze to the data. It is natural
to try to explain the hierarchical structure of these
parameters in terms of (gauged) chiral flavour symmetries
beyond those of the SM group. In section \ref{sec:chiral} we
report on recent developments in models which generate
realistic large mass ratios, via selection rules due to such
new approximately conserved chiral flavour quantum numbers.
The aim of such models is to reproduce all the fermion masses
and mixing angles within factors of order unity. Section
\ref{sec:con} contains our concluding remarks.

\section{Dynamical Determination of the Top Quark Mass}\label{sec:top}

We now consider some attempts to derive a fermion mass
or mass relation from some dynamical or theoretical principle.
Some time ago, Veltman \cite{veltman} suggested that
the fermionic and bosonic SM quadratic divergences should
cancel, in order that the ultra-violet cut-off for the theory
could be very high. It is, of course, one of the attractions
of the MSSM that such a cancellation is automatic in
supersymmetric (SUSY) theories. The Veltman condition for the
cancellation of quadratic divergences at one loop in the SM
gives the relation:
\begin{equation}
\sum_{leptons}m_l^2 + \sum_{quarks}m_q^2 =
\frac{3}{2} m_W^2 + \frac{3}{4} m_Z^2 + \frac{3}{4} m_H^2
\end{equation}
where the summation is over colour and flavour. It is not
clear at what scale \cite{jack} this relation should be
valid; but if applied at the electroweak scale it gives
a SM Higgs particle mass of $m_H \simeq 330$ GeV.

More recently, in a toy model, Nambu \cite{nambu} combined
the Veltman condition for the cancellation of quadratic
divergences with the idea that the vacuum energy density be
minimised with respect to the Yukawa couplings, keeping
all the other parameters fixed. This interesting idea
of making the Yukawa couplings dynamical, constrained by the Veltman
condition, naturally generates one Yukawa coupling
(identified with the top quark) much larger than
all the other ones \cite{nambu,bindud}.

The Nambu model immediately raises the question of
whether there is a
physical justification for treating the Yukawa couplings
as dynamical variables rather than numerical parameters.
There is a possible answer to this question in superstring
models, in which the Yukawa couplings depend on the vacuum
expectation values (VEVs) of some gauge singlet scalar
fields called moduli \cite{bindud,kounnas,dudas}.
These moduli fields, $T_{\alpha}$, which
parameterise the size and shape of the six-dimensional
compactified space, correspond to flat directions of
the scalar potential in
the effective four-dimensional supergravity theory.
Thus their VEVs are determined by quantum corrections,
with the possibility that some survive to the electroweak
scale. In this case the low energy MSSM effective potential
should be minimised with respect to these moduli, on which in
turn the Yukawa couplings depend. In other words the effective
potential should be minimised with respect to the
Yukawa couplings, possibly subjected to constraints
depending on the number of moduli fields remaining at
low energy. This minimisation tends to drive
the top quark mass to its largest allowed value and
hence close to its infra-red
quasi-fixed point value. Indeed, as we shall see,
this MSSM fixed point prediction for the top quark
is a common feature of many recent models
of quark masses.

In fact we expect, on rather general grounds, that all
coupling constants become dynamical in quantum gravity
due to the non-local effects of baby universes \cite{baby}.
As discussed in Holger Nielsen's talk here, we believe that
the mild form of non-locality in quantum gravity,
respecting reparameterisation invariance \cite{book},
leads to the realisation in Nature of the ``multiple
point criticality principle''. According to this
principle, Nature should choose coupling constant
values such that the vacuum can exist in
degenerate phases. In subsection \ref{sub:crit}
we apply this principle to the SM and obtain
predictions for the top quark and Higgs boson masses
\cite{smtop}, requiring the
VEVs in two degenerate SM vacua to differ by
an amount of order the Planck mass.

It has been known for some time \cite{maiani,sher} that
the self-consistency of the pure SM up to some
physical cut-off scale $\Lambda$ imposes constraints
on the top quark and Higgs boson masses.
The first constraint is the so-called triviality bound:
the running Higgs coupling constant $\lambda(\mu)$
should not develop a Landau pole for $\mu < \Lambda$.
The second is the vacuum stability bound:
the running Higgs coupling constant $\lambda(\mu)$
should not become negative leading to the instability of
the usual SM vacuum.
%
\begin{figure}
\leavevmode
\centerline{
\psfig{file=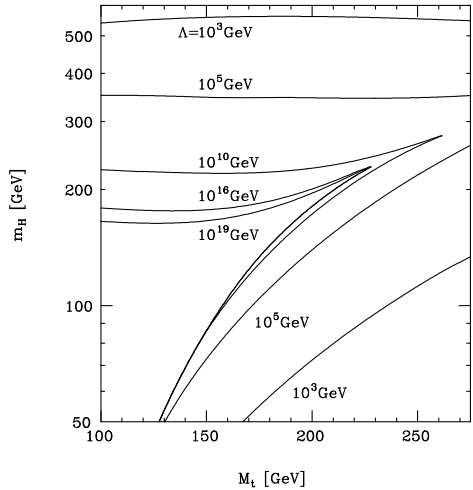,width=10.7cm,angle=90}
}
\vspace{-0.6cm}
\caption{SM bounds in the ($M_t$,$m_H$) plane ,
for  various values of $\Lambda$, the scale at which
new physics enters.}
\label{fig:Maiani}
\end{figure}
These bounds are illustrated \cite{zwirner} in
figure \ref{fig:Maiani}, where the combined triviality
and vacuum stability bounds for the SM are shown for different
values of the high energy cut-off $\Lambda$ . The allowed region
is the area around the origin bounded by the co-ordinate
axes and the solid curve labelled by the appropriate value
of $\Lambda$. In the following we shall be interested in
large cut-off scales $\Lambda \simeq 10^{15}-10^{19}$ GeV,
corresponding to the grand unified (GUT)
or Planck scale. The upper part of each
curve corresponds to the triviality bound. The lower part of
each curve coincides with the vacuum stability bound and the
point in the top right hand corner, where it meets the
triviality bound curve, is the quasi-fixed infra-red fixed point
for that value of $\Lambda$. The vacuum stability curve will
be important for the discussion of the SM criticality
prediction of the top quark and Higgs boson masses below. Before
this however, we will discuss their quasi-fixed point values
in the SM and the MSSM.

\subsection{Infrared Fixed Point Predictions}\label{sub:fixpt}

The idea that some of the properties of the quark-lepton mass spectrum
might be determined dynamically, as infrared fixed point values of the
renormalisation group equations (RGEs) for the Yukawa coupling constants,
was first considered \cite{fn1} some time ago.
It was pointed out that the three generation fermion
mass hierarchy does not develop naturally out of the general structure of
the RGEs. However it was soon realised \cite{pendleton} that
the top quark mass might correspond to
a fixed point value of the SM RGEs,
predicting approximately \mbox{$m_{t} \simeq 100$ GeV}
\cite{pendleton,marciano}.
In practice one finds that such an infrared fixed point
behaviour of the running top quark Yukawa coupling
constant $g_{t}(\mu)$ does not generically
set in until \mbox{$\mu < 1$ GeV},
where the QCD coupling constant $g_{3}(\mu)$
varies rapidly. The scale relevant for the physical top quark mass prediction
is of course $\mu = m_{t}$; at this scale $g_{3}(\mu)$ is slowly varying and
there is an effective infrared stable quasifixed point
(which would be an exact
fixed point if $g_{3}(\mu)$ were constant) behaviour \cite{hill}.

The quasifixed point prediction of the top quark mass
is based on two assumptions:
(a) the perturbative SM is valid up to
some high (e.~g.~GUT or Planck) energy scale
$\Lambda \simeq 10^{15} - 10^{19}$ GeV, and
(b) the top Yukawa coupling
constant is large at the high scale $g_{t}(\Lambda) \gsim 1$.
Neglecting the lighter quark masses and mixings, which is
a good approxmation, the SM one loop RGE for the top quark
Yukawa coupling $g_t(\mu) =
\sqrt{2} m_t(\mu)/\langle\phi_{WS}\rangle$ is:
\begin{equation}
16\pi^2\frac{dg_t}{d\ln\mu} = g_t\left(\frac{9}{2}g_t^2 - 8g_3^2
- \frac{9}{4}g_2^2 - \frac{17}{12}g_1^2\right)
\end{equation}
The gauge coupling constants $g_i(\mu)$ satisfy the RGEs:
\begin{equation}
16\pi^2\frac{dg_i}{d\ln\mu} = b_ig_i^3 ; \qquad \qquad \rm{where} \qquad
b_1 = \frac{41}{6}, \quad b_2 = -\frac{19}{6}, \quad b_3 =-7
\qquad \rm{in \; the \; SM.}
\label{rgegauge}
\end{equation}

The nonlinearity of the RGEs
then strongly focuses $g_{t}(\mu)$ at the
electroweak scale to its quasifixed point value. We note that
while there is a rapid convergence to the top Yukawa coupling fixed
point value from above, the approach from below is much more gradual.
The RGE for the Higgs self-coupling $\lambda(\mu) =
m_H^2(\mu)/\langle\phi_{WS}\rangle^2$
\begin{equation}
16\pi^2\frac{d\lambda}{d\ln\mu} =12\lambda^2 +
3\left(4g_t^2 - 3g_2^2 - g_1^2\right)\lambda +
\frac{9}{4}g_2^4 + \frac{3}{2}g_2^2g_1^2 + \frac{3}{4}g_1^4 - 12g_t^4
\label{rgelam}
\end{equation}
similarly focuses $\lambda(\mu)$ towards a
quasifixed point value, leading to the SM fixed point predictions \cite{hill}
for the running top quark and Higgs masses:
\begin{equation}
m_{t} \simeq 225\ \mbox{GeV} \quad m_{H} \simeq 250\ \mbox{GeV}
\end{equation}
Unfortunately these predictions are inconsistent with the CDF and D0
results \cite{CDF}, which require a
running top mass \mbox{$m_{t} \simeq 170 \pm 12$ GeV}.
Note that the running
quark mass $m_{q}(\mu)$ is related to the  physical
or pole quark mass $M_{q}$,
defined as the location of the pole in the quark propagator, by
\begin{equation}
M_{q} = m_{t}(M_{q})\left(1 + 4\alpha_{3}(M_{q})/3\pi\right)
\end{equation}
at the one loop QCD level. The top quark pole mass is thus
5-6\% larger than its running mass and we have taken \cite{CDF}
$M_t = 180 \pm 12$ GeV. Consideration of higher order effects
and the definition of the Higgs boson pole mass introduce many
complicated issues \cite{drtjones,shervs,isidori,casas,willey}
involving the SM renormalisation procedure.

\begin{figure}[b]
\leavevmode
\centerline{
\epsfxsize=6.75cm
\epsfbox{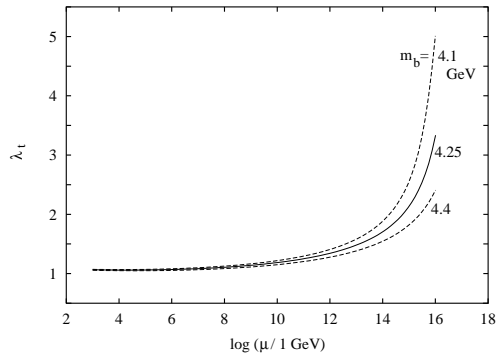}
}
\caption{The rapid convergence from above of the top Yukawa coupling constant
$\lambda_t \equiv g_t$ to the MSSM infra-red quasi-fixed point value
as $\mu \rightarrow m_t$.}
\label{fig:bargfp}
\end{figure}

There are two interesting modifications to the fixed point top mass
prediction in the MSSM with
supersymmetry breaking at the electroweak scale or TeV scale:
\begin{itemize}
\item
The introduction of the supersymmetric partners of the SM particles in the
RGE for the Yukawa and gauge coupling constants leads to a 15\% reduction in
the fixed point value of $g_{t}(m_{t})$ \cite{bagger,dimhallrab}.
\item
There are two Higgs doublets in the MSSM and the
ratio of Higgs vacuum values, $\tan \beta = v_{2}/v_{1}$, is a free parameter;
the top quark couples to $v_{2}$ and so $m_{t}$ is proportional to
\mbox{$v_{2} = (174\ \mbox{Gev})\sin\beta$}.
\end{itemize}
The RGE for the top quark Yukawa coupling constant
in the MSSM becomes
\begin{equation}
16\pi^2\frac{dg_t}{d\ln\mu} = g_t\left(6g_t^2 + g_b^2 -
\frac{16}{3}g_3^2 - 3g_2^2 - \frac{13}{9}g_1^2\right)
\end{equation}
For $\tan \beta$ of order unity, the Yukawa couplings
of the bottom quark $g_b(\mu)$ and the tau lepton $g_{\tau}(\mu)$
can still be neglected.
The RGEs for the gauge coupling constants, eq. (\ref{rgegauge}),
are used with the supersymmetric values:
\begin{equation}
b_1 = 11, \qquad b_2 = 1, \qquad b_3 = -3
\end{equation}
The approach from above of $g_t(\mu)$ to its MSSM infrared
fixed point value value  of
approximately 1.1 is shown \cite{barger} in figure \ref{fig:bargfp}.
The corresponding MSSM fixed point prediction for the
top quark pole mass is
\cite{barger}:
\begin{equation}
m_{t}(m_{t}) \simeq (200\ \mbox{GeV})\sin\beta
\label{mssmfp}
\end{equation}
which is remarkably close to the LEP and CDF results for
\mbox{$\tan\beta > 1$}.

\begin{figure}
\leavevmode
\centerline{
\epsfxsize=6.75cm
\epsfbox{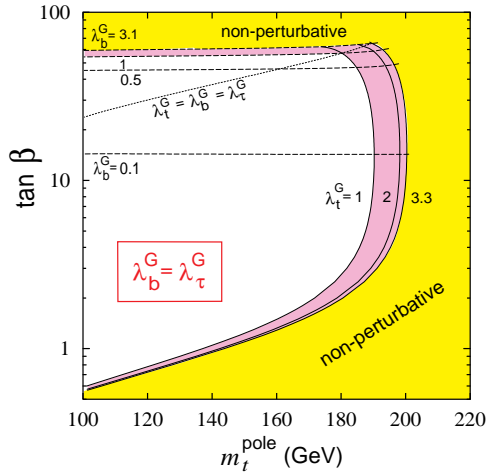}
}
\caption{The top quark MSSM infrared fixed point region, corresponding
to large values of its Yukawa coupling constant at the SUSY-GUT scale
($\lambda_t^G \equiv g_t^G \gsim 1$). Also shown are values of
the bottom quark and tau lepton Yukawa couplings chosen to be
equal at the SUSY-GUT scale.}
\label{fig:barg2fp}
\end{figure}

The above quasifixed point value is of course also
the upper bound on the top mass
in the MSSM, assuming perturbation theory is valid in the desert up to the
SUSY-GUT scale. It then follows  that the experimental
evidence for a large top mass requires
\mbox{$\tan\beta > 1$}. We note that the minimal SU(5) SUSY-GUT symmetry
relation between the bottom quark and tau lepton Yukawa coupling constants,
\mbox{$g_{b}(\Lambda) = g_{\tau}(\Lambda)$}, is also only satisfied
phenomenologically
if the top quark Yukawa coupling is close to its infrared quasifixed point
value,
so that it contributes significantly to the running of \mbox{$g_{b}(\mu)$} and
reduces the predicted value of $m_{b}(m_{b})$. In the SM the contribution
of the top quark Yukawa coupling has the opposite sign and the SU(5) GUT
prediction for $m_{b}(m_{b})$ fails, as it is then phenomenologically too
large.

The MSSM infrared fixed point value for the top quark pole mass is shown
\cite{barger} in figure \ref{fig:barg2fp} as a function of $\tan\beta$.
For large $\tan\beta$ it is possible to have a bottom quark Yukawa coupling
satisfying \mbox{$g_{b}(\Lambda) \gsim 1$} which then approaches an infrared
quasifixed point and is no longer negligible in the RGE for $g_{t}(\mu)$.
Indeed with
$\tan\beta \simeq m_{t}(m_{t})/m_{b}(m_{t}) \simeq 60$,
we can trade the mystery of the top to bottom quark mass ratio
for that of a
hierarchy of vacuum expectation values, \mbox{$v_{2}/v_{1} \simeq
m_{t}(m_{t})/m_{b}(m_{t})$},
 and have all the third generation Yukawa coupling constants large:
\begin{equation}
g_{t}(\Lambda) \gsim 1 \quad g_{b}(\Lambda) \gsim 1 \quad
g_{\tau}(\Lambda) \gsim 1
\label{tbtaufp}
\end{equation}
Then $m_{t}$, $m_{b}$ and \mbox{$R = m_{b}/m_{\tau}$} all approach infrared
quasifixed point
values compatible with experiment \cite{fkm}. This large $\tan\beta$
scenario is consistent with the idea of Yukawa unification \cite{anan}:
\begin{equation}
g_{t}(\Lambda) = g_{b}(\Lambda) = g_{\tau}(\Lambda) = g_{G}
\label{yukun}
\end{equation}
as occurs in the SO(10) SUSY-GUT model with
the two MSSM Higgs doublets in a single {\bf 10} irreducible representation
and $g_{G} \gsim 1$ ensures fixed point behaviour.
However it should be noted that the equality in
eq.~(\ref{yukun}) is not necessary to obtain the fixed point values.
For example the principles of finiteness and the reduction
of couplings \cite{zoupanos} can be used to relate
the Yukawa couplings to the SUSY-GUT coupling constant.
In SU(5) finite unified theories  the relationship is
\mbox{$g_{t}^{2}(M_{X}) = 4g_{b}^{2}(M_{X})/3$ = $\cal O$(1)}, giving a
fixed point prediction similar to eq.~(\ref{mssmfp}).
In fact one does not need a symmetry assumption at all, since the weaker
assumption of large third generation Yukawa couplings, eq.~(\ref{tbtaufp}),
is sufficient for the fixed point dynamics to predict \cite{fkm}
the running masses
$m_{t} \simeq 180 \ \mbox{GeV}$, $m_{b} \simeq 4.1 \ \mbox{GeV}$ and
$m_{\tau} \simeq 1.8 \ \mbox{GeV}$ in the large $\tan\beta$ scenario.
Also the lightest Higgs particle mass is predicted to be
$m_{h^0} \simeq 120 \
\mbox{GeV}$ (for a top squark mass of order  \mbox{1 TeV}).
We also note that in superstring models tree level Yukawa couplings
are proportional to the unified gauge coupling constant $g_G$,
with a constant of proportionality either of order unity or
exponentially suppressed. If the third generation Yukawa couplings
have no residual moduli dependence at the electroweak scale, this
superunification condition on the top quark Yukawa coupling
$g_t(\Lambda)$ (also $g_b(\Lambda)$ and $g_{\tau}(\Lambda)$ for
large $\tan\beta$) ensures that it is attracted towards its infrared
fixed point value by its RGE anyway.

\begin{figure}
\leavevmode
\vspace{-2.5cm}
\centerline{
\epsfxsize=13cm
\epsfbox{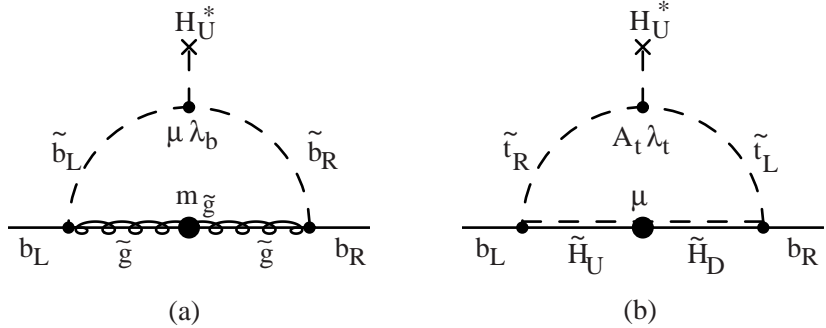}
}
\vspace{-9.7cm}
\caption{Leading one-loop MSSM contributions to
the b quark mass.}
\label{fig:bloops}
\end{figure}

The origin of the large value of \mbox{$\tan\beta$} is of course a puzzle,
which
must be solved before the large \mbox{$\tan\beta$} scenario can be said to
explain the large \mbox{$m_{t}/m_{b}$} ratio. It is possible to introduce
approximate symmetries \cite{anderson,hall} of the Higgs potential which
ensure a hierarchy of vacuum expectation values - a Peccei-Quinn symmetry and
a continuous $\cal R$ symmetry have been used. The Peccei-Quinn
symmetry essentially forbids the $\mu$-term in the superpotential,
while the $\cal R$ symmetry requires the vanishing of
gaugino masses $M_{1/2}$, of the SUSY-breaking trilinear scalar
couplings $A_i$ and of the bilinear SUSY-breaking Higgs coupling B.
However these symmetries seem to be inconsistent with the popular
scenario of universal soft SUSY breaking mass parameters at the unification
scale and radiative electroweak symmetry breaking \cite{carena}.

Also, in the large $\tan\beta$ scenario, SUSY radiative corrections to $m_{b}$
are generically large: the bottom quark mass gets a contribution proportional
to $v_{2}$ from some one-loop diagrams,
whereas its tree level mass is proportional to
$v_{1} = v_{2}/\tan\beta$.
The two dominant diagrams correspond to a bottom squark-gluino loop
and a top squark-charged Higgsino loop respectively
\cite{anderson,hall}, as shown in figure \ref{fig:bloops}.
Consequently these loop diagrams give a
fractional correction \mbox{$\delta m_{b}/m_{b}$} to the bottom quark mass
proportional to $\tan\beta$ and generically of order unity
\cite{hall,carena}. The presence of
the above-mentioned Peccei-Quinn and $\cal R$ symmetries and the associated
hierarchical SUSY spectrum (with the third generation
squarks and sleptons much heavier than
the gauginos and Higgsinos) would protect $m_{b}$ from large radiative
corrections, by providing a suppression factor in the loop diagrams and
giving \mbox{$\delta m_{b}/m_{b} \ll 1$}.
However, in the absence of
experimental information on the superpartner spectrum,
precise predictions of the
third generation quark-lepton masses in the large $\tan\beta$ scenario
are, unfortunately, not possible.

\subsection{Standard Model Criticality Predictions}\label{sub:crit}
\begin{figure}
\leavevmode
\vspace{-0.5cm}
\centerline{
\psfig{file=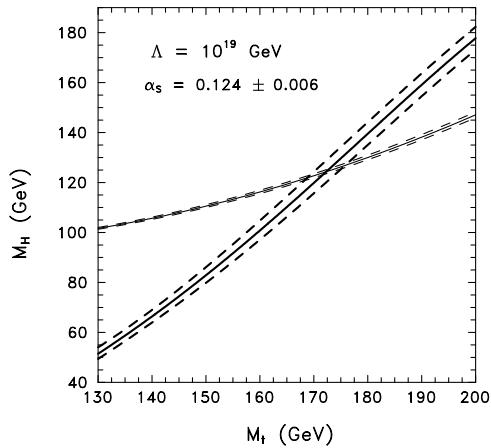,width=7cm,%
bbllx=95pt,bblly=130pt,bburx=510pt,bbury=555pt,%
angle=270,clip=}
}
\caption{Diagonal (thick) lines: SM vacuum stability curve
for $\Lambda = 10^{19}$ GeV and $\alpha_s = 0.124$ (solid line),
$\alpha_s = 0.118$ (upper dashed line), $\alpha_s = 0.130$
(lower dashed line). Transverse (thin) lines: MSSM upper
bounds on the Higgs pole mass $M_H$
for a SUSY breaking scale of 1 TeV and
$\alpha_s$ as in the diagonal lines.}
\label{fig:vacstab}
\end{figure}

Here we consider the idea \cite{glasgowbrioni},
discussed in more detail by Holger Nielsen at this meeting, that
Nature should choose coupling constant
values such that several ``phases'' can coexist, in a very similar way to
the stable coexistence of ice, water and vapour (in a thermos
flask for example) in a mixture with fixed energy and number of molecules.
The application of this so-called multiple point criticality principle
to the determination of the top quark Yukawa coupling constant requires
the SM (renormalisation group improved) effective Higgs potential to have
coexisting vacua, which means degenerate minima:
\begin{equation}
V_{\rm{eff}}(\phi_{\rm{min}\; 1}) = V_{\rm{eff}}(\phi_{\rm{min} \; 2})
\label{eqdeg}
\end{equation}
This condition really means that the vacuum in which we live
is barely stable; we just lie on the vacuum stability curve
mentioned above and shown \cite{casas}
in figure \ref{fig:vacstab} for a cut-off
$\Lambda = 10^{19}$ GeV.
The important point
for us, in the analogy of the ice, water and vapour system, is that the
choice of the fixed extensive variables, such as energy, the number of
moles and the volume, can very easily be such that a mixture must occur.
In that case then the temperature and pressure (i.\ e.\ the intensive
quantities) take very specific values, namely the values at the triple
point, without any finetuning. We stress that this phenomenon of thus
getting specific intensive quantitities is only {\em likely} to happen
for stongly first order phase transitions, for which the interval of
values for the extensive variables that can only be realised as an
inhomogeneous mixture of phases is rather large.

In the analogy considered here, the SM top quark Yukawa
coupling and the Higgs self coupling correspond to intensive quantities
like temperature and pressure. If these couplings are to be determined
by the criticality condition, the two phases corresponding to
the two effective
Higgs field potential minima should have some ``extensive quantity'',
such as $\int d^4x |\phi(x)|^2 $, deviating ``strongly'' from
phase to phase. If, as we shall assume, Planck units reflect the fundamental
physics it would be natural to interpret this strongly first order
transition requirement to mean that, in Planck units, the extensive
variable densities  $\frac{\int d^4x |\phi(x)|^2}{  \int d^4x }$ = $<|\phi|^2>$
for the two vacua should differ by a quantity of order unity.
Phenomenologically we know that
for the vacuum 1 in which we live,
$<|\phi|^2>_{\rm{vacuum}\; 1}$ =  (246 GeV)$^2$
and thus we should really expect the Higgs field
in the other phase just to be of Planck order of magnitude:
\begin{equation}
<|\phi|^2>_{\rm{vacuum}\; 2} \qquad = \qquad
{\cal O}(\rm{M}_{\rm{Planck}}^2)
\label{eqstrong}
\end{equation}
This strong first orderness condition, eq. (\ref{eqstrong}), and
the degeneracy condition, eq. (\ref{eqdeg}), are our two
crucial assumptions.

\begin{figure}[t]
\leavevmode
\centerline{
\epsfxsize=6.75cm
\epsfbox{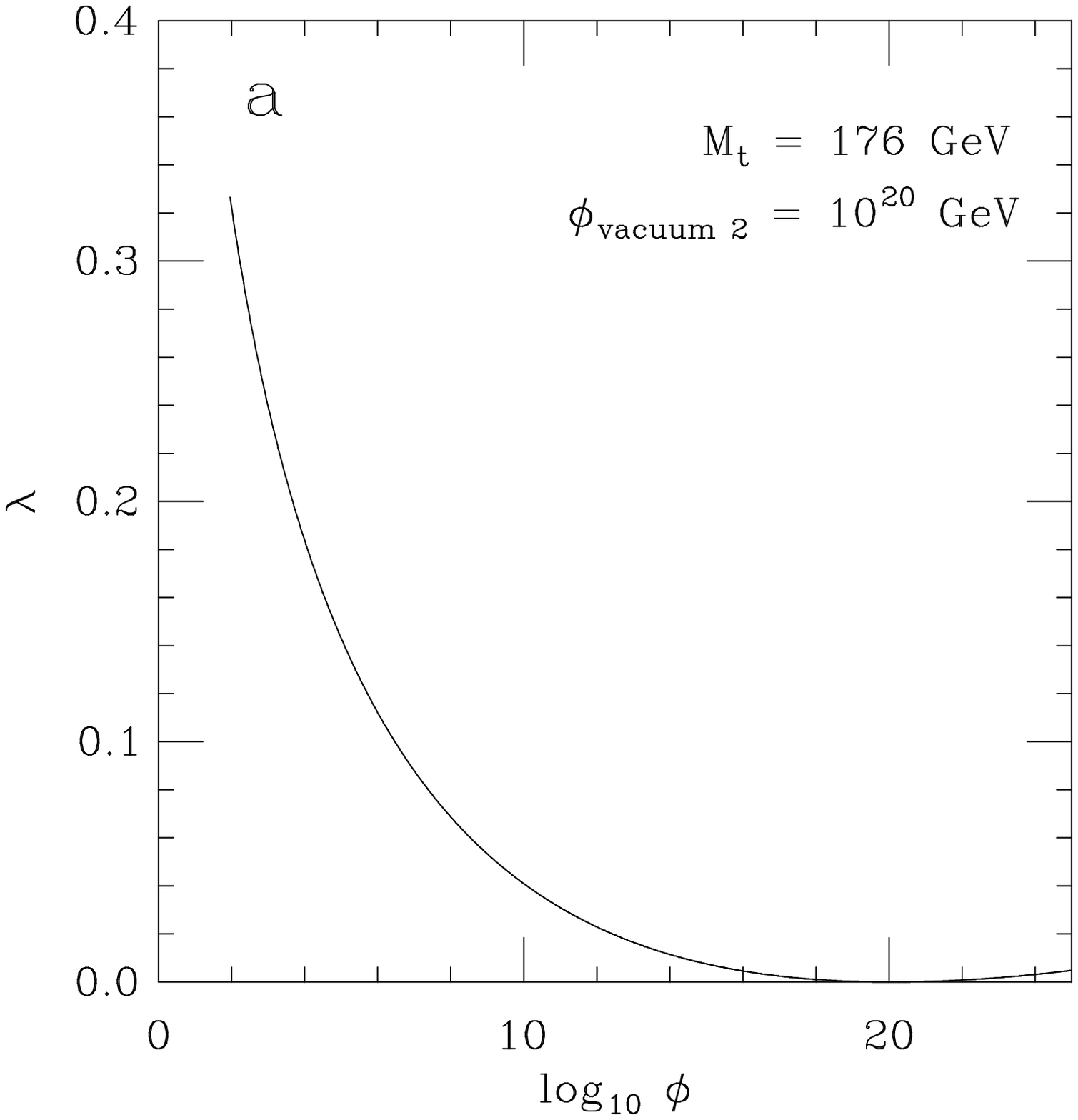}
\epsfxsize=6.75cm
\epsfbox{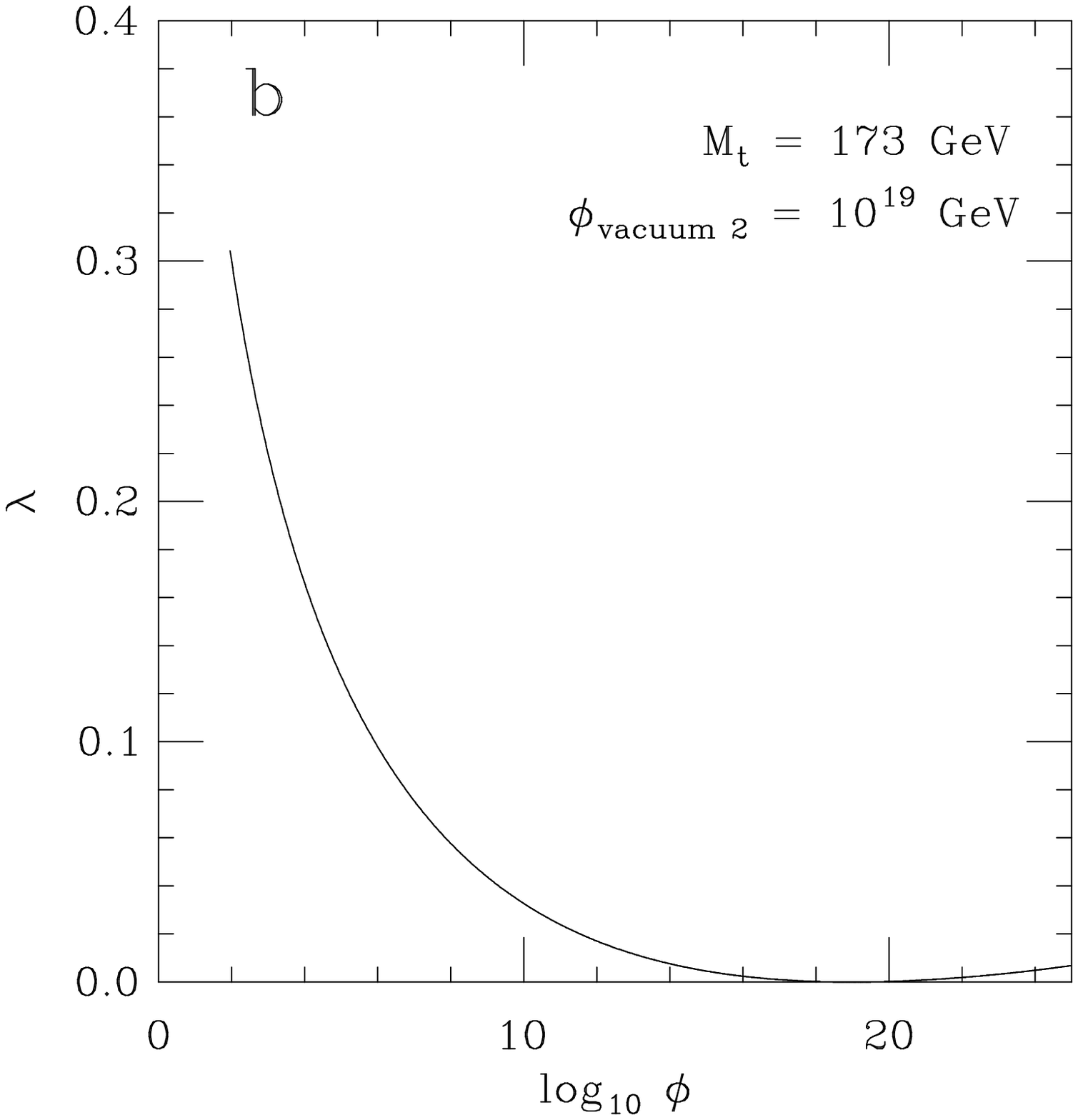}
}
\caption{Plot of $\lambda$ as a function of the scale of the Higgs field
$\phi$ for degenerate vacua with the second Higgs VEV at the scale
(a) $\phi_{\rm{vacuum} \; 2} = 10^{20}$ GeV and
(b) $\phi_{\rm{vacuum} \; 2} = 10^{19}$ GeV.
We formally apply the second order SM renormalisation group equations up to
a scale of $10^{25}$ GeV.}
\label{fig:figure}
\end{figure}

In vacuum 2 the $\phi^4$ term
will a priori strongly dominate the $ \phi^2$
term. So we
basically get the degeneracy to mean that, at the vacuum 2 minimum,
the effective coefficient $\lambda(\phi_{\rm{vacuum}\; 2})$ must be
zero with high accuracy. At the same $\phi$-value the derivative
of the renormalisation group improved
effective potential $V_{\rm{eff}}(\phi)$ should be zero because
it has a minimum there.
Thus at the second minimum the beta-function
$\beta_{\lambda}$ vanishes as well as $\lambda(\phi)$, which
gives to leading order, setting $\lambda = 0$ in
eq. (\ref{rgelam}), the condition:
\begin{equation}
\frac{9}{4}g_2^4 + \frac{3}{2}g_2^2g_1^2 + \frac{3}{4}g_1^4 - 12g_t^4 = 0
\end{equation}
We use the renormalisation group to relate the couplings
at the scale of vacuum 2, i.~e.~at $\mu= \phi_{\rm{vacuum}\; 2}$,
to their values
at the scale of the masses themselves, or roughly at the electroweak scale
$\mu \approx <\phi>_{\rm{vacuum} \; 1}$.
Figure \ref{fig:figure}
shows the running $\lambda(\phi)$ as a function of $\log(\phi)$ computed
for two values of $\phi_{vacuum\; 2}$ (where we impose the conditions
$\beta_{\lambda} = \lambda = 0$) near the Planck scale
$M_{\rm{Planck}} \simeq 10^{19}$ GeV.
Combining the
uncertainty from the Planck scale only being known in order of
magnitude and the QCD fine structure constant
$\alpha_s(M_Z) = 0.117 \pm 0.006$ uncertainty
with the calculational uncertainty
\cite{drtjones,shervs,isidori,casas,willey},
we get our predicted combination of top and Higgs pole masses:
\begin{equation}
M_{t} = 173 \pm 4\ \mbox{GeV} \quad M_{H} = 135 \pm 9\ \mbox{GeV}.
\end{equation}

\section{Mass Matrix Ans\"{a}tze}\label{sec:ansatz}
The best known ansatz for the quark mass matrices
is due to Fritzsch \cite{fritzsch}:
\begin{equation}
M_U =\pmatrix{0  		& C   		& 0\cr
		      C  		& 0   		& B\cr
		      0  		& B   		& A\cr}
\qquad
M_D =\pmatrix{0  		& C^\prime  & 0\cr
		      C^\prime 	& 0   		& B^\prime\cr
		      0  		& B^\prime  & A^\prime\cr}
\end{equation}
It contains 6 complex parameters A, B, C, $A^\prime$, $B^\prime$ and
$C^\prime$, where it is necessary to {\em assume}:
\begin{equation}
|A| \gg |B| \gg |C|, \qquad |A^\prime| \gg |B^\prime| \gg |C^\prime|
\end{equation}
in order to obtain a good fermion mass hierarchy.
Four of the phases can be rotated away by redefining the phases of
the quark fields, leaving just 8 real parameters (the magnitudes of
A, B, C, $A^\prime$, $B^\prime$ and $C^\prime$ and two phases
$\phi_{1}$ and $\phi_{2}$) to reproduce 6 quark masses and
4 angles parameterising $V_{CKM}$. There are thus two relationships predicted
by the Fritzsch ansatz:
\begin{equation}
|V_{us}| \simeq
\left| \sqrt{\frac{m_{d}}{m_{s}}} -
e^{-i\phi_{1}}\sqrt{\frac{m_{u}}{m_{c}}} \right|,
\qquad \qquad |V_{cb}| \simeq
\left| \sqrt{\frac{m_{s}}{m_{b}}} -
e^{-i\phi_{2}}\sqrt{\frac{m_{c}}{m_{t}}} \right|
\label{fritzsch1}
\end{equation}
The first prediction is a generalised version of the relation
$\theta_c\simeq\sqrt{\frac{m_d}{m_s}}$ for the Cabibbo angle,
which originally motivated the ansatz
and is well satisfied experimentally. However the second relationship
cannot be satisfied with a heavy top quark mass
\mbox{$m_{t} > 100$ GeV} \cite{gilman} and
the original Fritzsch ansatz is excluded by the data.
Consistency with experiment can, for example, be restored by
introducing a non-zero 22 mass matrix element \cite{xing},
or by breaking the symmetry of the mass matrices and
making the (2,3) and (3,2) elements of $M_U$ unequal \cite{dutta}.

It is now common to make ans\"{a}tze
incorporating relationships between the fermion mass
parameters at the GUT or the Planck scale.
We have already mentioned the best known result: the simple SU(5) relation
$m_{b}(\Lambda) = m_{\tau}(\Lambda)$ which is
satisfied in SUSY-GUTs provided the
top quark mass is near to its quasifixed point value \cite{dimhallrab,arason}.
However the corresponding relations for the first two generations are not
satisfied, as they predict for example
\begin{equation}
m_{d}/m_{s} = m_{e}/m_{\mu}
\label{eq:massratio}
\end{equation}
which fails phenomenologically by an order of magnitude. This led Georgi and
Jarlskog \cite{georgijarlskog,harvey} to postulate the mass relations
$m_{b}(M_X) = m_{\tau}(M_{X})$, $m_{s}(M_X) = m_{\mu}(M_{X})$/3 and
$m_{d}(M_X) = 3m_{e}(M_{X})$ at the GUT scale.
Dimopoulos, Hall and Raby \cite{dimhallrab} revived these relations
in the context of an SO(10) SUSY-GUT, combining the Fritzsch form for the up
quark mass matrix \mbox{$M_{U} = Y_{u}v_{2}$} with the Georgi-Jarlskog form
for the down quark and charged lepton mass matrices
\mbox{$M_{D} = Y_{d}v_{1}$} and \mbox{$M_{L} = Y_{l}v_{1}$}:

\begin{equation}
Y_u =\pmatrix{0  		 & C   				& 0\cr
		      C  		 & 0   				& B\cr
		      0  		 & B   				& A\cr}
\quad
Y_d =\pmatrix{0  		 & Fe^{i\phi}		& 0\cr
		      F^{-i\phi} & E   				& 0\cr
		      0  		 & 0				& D\cr}
\quad
Y_l =\pmatrix{0  		& F   				& 0\cr
		      F  		& -3E  				& 0\cr
		      0  		& 0   				& D\cr}
\label{eq:dhransatz}
\end{equation}
The  phase freedom in the definition of the fermion fields has been
used to make the parameters A, B, C, D, E and F real and we have again to
assume:
$|A| \gg |B| \gg |C|$ and $|D| \gg |E| \gg |F|$.
Thus there are 7 free parameters in the Yukawa coupling ansatz and
$\tan\beta$ available to fit 13 observables. Using the RGEs from the
SUSY-GUT scale to the electroweak scale, this ansatz gives 5 predictions
which, with the possible exception of $V_{cb}$ discussed below,
are in agreement with data for
\mbox{$1 < \tan\beta < 60$} \cite{dimhallrab,bargander}. The simple
SU(5) prediction, eq.~(\ref{eq:massratio}), is replaced
by the successful mass ratio prediction
\begin{equation}
(m_{d}/m_{s}) (1 - m_{d}/m_{s})^{-2} =
9 (m_{e}/m_{\mu}) (1 - m_{e}/m_{\mu})^{-2}
\end{equation}
Since the down quark matrix $Y_{d}$ is diagonal in
the two heaviest generations,
one of the SUSY-GUT scale predictions is
\mbox{$V_{cb} \simeq \sqrt{\frac{m_{c}}{m_{t}}}$} \cite{harvey}.
Fits give $m_{t}$ close to its fixed point and the large top
Yukawa coupling causes  $V_{cb}$ to run
between the GUT and electroweak scales to a somewhat lower value. Nonetheless
the fits still tend to make $V_{cb}$ greater than
the experimental value $0.040 \pm 0.005$ \cite{pdg}.
A fit satisfying Yukawa unification is obtained by setting
\mbox{$A = D$} and \mbox{$\tan\beta \simeq 60$},
for which $|V_{cb}| \ge 0.052$. This fit is of course subject to
uncertainties due to the possibly large SUSY radiative corrections to $m_{b}$
mentioned in the previous section and to other elements of the
down quark mass matrix \cite{blazek}.

It is in fact possible to make a
systematic analysis \cite{texture} of {\em symmetric} quark mass matrices
with 5 or 6 ``texture'' zeros (counting pairs
of off-diagonal zeros as one zero
due to the symmetric structure assumed). The analysis
is simply extended to include leptons by using
the Georgi-Jarlskog ansatz: all the elements of
$Y_l$ are taken the same as for $Y_d$ except the
(2,2) element which is multiplied by the factor 3.
The assumed  symmetry of the Yukawa coupling
matrices $Y_u$, $Y_d$ and $Y_l$ in generation
space does not of course necessarily hold in the SM.
However for unified models in which each generation
appears as a single representation, such as SO(10),
this symmetry is quite natural. Pragmatically this
symmetry makes the analysis more tractable. It is also
convenient to redefine the  phases of the quark and lepton
fields in such a way that the $3\times 3$ symmetric
Yukawa matrices are transformed into hermitean matrices.
There are just 6 possible forms of symmetric Yukawa matrices
with a hierarchy of three non-zero eigenvalues and the
maximum number (three) of texture zeros.
A survey of all six and five texture zero structures
for the quark mass matrices, applied at the SUSY-GUT scale,
reveals a total of 5 ans\"{a}tze, each with five
texture zeros, consistent with experiment.
The corresponding quark Yukawa matrices
are given in table \ref{table:rrr}. Note that solution 2
with $E^{\prime} = 0$ reduces to the ansatz \cite{dimhallrab}
of eq. (\ref{eq:dhransatz}). This ansatz with six texture zeros
was excluded from table \ref{table:rrr} since it gives
large values for $V_{cb}$.

\begin{table}
\centering
\begin{tabular}{|c|c|c|c|c|c|} \hline
 & & & & & \\
Solution & 1 & 2 & 3 & 4 & 5 \\
 & & & & & \\
\hline
 & & & & & \\
${\bf Y}_u$ &
$\left(
\begin{array}{ccc}
0 & C & 0 \\
C & B & 0 \\
0 & 0 & A
\end{array}
\right)$ &
$\left(
\begin{array}{ccc}

0 & C & 0 \\
C & 0 & B \\
0 & B & A
\end{array}
\right)$ &
$\left(
\begin{array}{ccc}

0 & 0 & C \\
0 & B & 0 \\
C & 0 & A
\end{array}
\right)$ &
$\left(
\begin{array}{ccc}

0 & C & 0 \\
C & B & B^{\prime} \\
0 & B^{\prime} & A
\end{array}
\right)$ &
$\left(
\begin{array}{ccc}

0 & 0 & C \\
0 & B & B^{\prime} \\
C & B^{\prime} & A
\end{array}
\right)$
\\
 & & & & & \\
\hline
 & & & & & \\
${\bf Y}_d$ &
$\left(
\begin{array}{ccc}

0 & F & 0 \\
F^{\star} & E & E^{\prime} \\
0 & E^{\prime} & D
\end{array}
\right)$
&
$\left(
\begin{array}{ccc}

0 & F & 0 \\
F^{\star} & E  & E^{\prime} \\
0 & E^{\prime \star} & D
\end{array}
\right)$
&
$\left(
\begin{array}{ccc}

0 & F & 0 \\
F^{\star} & E & E^{\prime} \\
0 & E^{\prime } & D
\end{array}
\right)$
& $\left(
\begin{array}{ccc}

0 & F & 0 \\
F^{\star} & E & 0 \\
0 & 0 & D
\end{array}
\right)$
& $\left(
\begin{array}{ccc}

0 & F & 0 \\
F^{\star} & E & 0 \\
0 & 0 & D
\end{array}
\right)$
\\
  &  &  \\
\hline
\end{tabular}
\caption{Symmetric textures for the quark Yukawa matrices
at the SUSY-GUT scale, which are consistent with the measured values
of quark masses and mixing angles.}
\label{table:rrr}
\end{table}

The parameters in the Yukawa matrices of the solutions in
table \ref{table:rrr} have a hierarchical
structure similar to those in eq.~(\ref{eq:dhransatz}).
It is convenient \cite{texture} to reparameterise the Yukawa matrices in a
way that keeps track of the order of magnitude of the various
elements, as was done by Wolfenstein
\cite{wolfenstein} for the quark mixing matrix
in powers of the Cabibbo angle $\lambda \simeq 0.22$:
\begin{equation}
V_{CKM} =\pmatrix{1 - \lambda^2/2 & \lambda & A\lambda^3(\rho - i\eta) \cr
		    	 - \lambda & 1 -\lambda^2/2 &         A\lambda^2 \cr
	 A\lambda^3(1 - \rho -i\eta) & -A\lambda^2 &                1 \cr}
+ {\cal O}(\lambda^4)
\label{eq:wolf}
\end{equation}
For example solution 2 is well approximated by the following form:
\begin{equation}
Y_u =\pmatrix{   0      	& \lambda^6  & 0 \cr
		\lambda^6  	& 0 	     & \lambda^2  \cr
		 0 		& \lambda^2  & 1 \cr}
\qquad
Y_d = \pmatrix{   0 	   & 2\lambda^4 & 0 \cr
		2\lambda^4 & 2\lambda^3 & 2\lambda^3 \cr
			 0 & 2\lambda^3 & 1 \cr}
\label{eq:solution2}
\end{equation}
It is natural to interpret the small parameter $\lambda$ as a symmetry
breaking parameter for some approximate symmetry beyond those of
the Standard Model Group (SMG). The nature of
this symmetry is discussed in the next section.

The neutrino Majorana mass matrices generated by the see-saw mechanism in
many extensions of the SM naturally have the above type of symmetric texture.
Due to the hierarchical structure of their elements, there are two
qualitatively different types of eigenstate that can arise
\cite{fn2}. In the first case,
a neutrino can dominantly combine with its own antineutrino to form a
Majorana particle. The second case occurs when a neutrino combines
dominantly with an antineutrino, which is not the CP conjugate state,
to form a
2-component massive neutrino. For example the electron neutrino might combine
with the muon antineutrino. Such states naturally occur in pairs with
order of magnitude-wise degenerate masses. In the example given, the other
member of the pair of Majorana states would be formed by combining the muon
neutrino with the electron antineutrino. The hierarchical structure
which gives rise to this second case is of course ruled out phenomenologically
for the quark and charged lepton mass matrices, as none have a pair of states
with order of magnitude-wise degenerate masses. However, considering
two generations for simplicity, a neutrino mass matrix of the form
\begin{equation}
M_{\nu} \quad = \quad
\bordermatrix{	 &  \nu_1	&  \nu_2	\cr
		\overline{\nu}_1 &     0	&	  B		\cr
		\overline{\nu}_2 &	   B	&	  A		\cr}
\end{equation}
with the assumed hierarchy
\begin{equation}
|B| \gg |A|
\end{equation}
could be phenomenologically relevant. The mass eigenvalues are
$m_1 = B + A/2$ and $m_2 = B - A/2$, giving a neutrino mass squared difference
$\Delta m^2 = 2AB$, and the neutrino mixing angle is
$\theta \simeq \pi/4$ giving maximal mixing.
Maximal neutrino mixing, $\sin^2 2\theta \simeq 1$, provides a candidate
explanation for
(i) the atmospheric muon neutrino deficit with
$\Delta m^2 = 10^{-2} eV^2$ and $\nu_{\mu}$-$\nu_{\tau}$ oscillations, or
(ii) the solar neutrino problem with $\Delta m^2 = 10^{-10} eV^2$ and
$\nu_{e}$-$\nu_{\mu}$ vacuum oscillations.

\section{The Fermion Mass Hierarchy and Chiral Symmetry}\label{sec:chiral}
The fermion mass hierarchy can be expressed in terms of the Cabibbo angle
$\lambda \simeq 0.22$ as a small expansion parameter by the order
of magnitude values:
\begin{equation}
\frac{m_u}{m_t} = {\cal O}(\lambda^8), \quad
\frac{m_c}{m_t} = {\cal O}(\lambda^4), \quad
\frac{m_d}{m_b} = {\cal O}(\lambda^4), \quad
\frac{m_s}{m_b} = {\cal O}(\lambda^2), \quad
\frac{m_e}{m_{\tau}} = {\cal O}(\lambda^4 - \lambda^5), \quad
\frac{m_{\mu}}{m_{\tau}} = {\cal O}(\lambda^2)
\label{eq:massratios}
\end{equation}
at the high energy, GUT or Planck, scale.
It is natural to try to explain the occurrence of
these large mass ratios in terms of
selection rules due to approximate conservation laws.
The mass $m$ in the Dirac equation
$i\dslash \psi_L = m \psi_R$ essentially represents
a transition amplitude between
a left-handed fermion component $\psi_L$ and its right-handed partner
$\psi_R$.
If $\psi_L$ and $\psi_R$ have different quantum numbers, i.~e.~belong to
inequivalent irreducible representations (IRs)
of a symmetry group $G$ ($G$ is
then called a {\em chiral\/} symmetry),
the mass term is forbidden in the
limit of exact $G$ symmetry and they represent
two massless Weyl particles. $G$
thus ``protects'' the fermion from gaining a mass.
Note that this is exactly the situation for all the SM
fermions, which are mass-protected by $SU(2)_L
\times U(1)_Y$ (but not by $SU(3)_c$). The $SU(2)_L
\times U(1)_Y$ symmetry is spontaneously broken
and the  SM fermions gain masses
suppressed relative to the presumed fundamental
(GUT or Planck) mass scale $M$ by
the symmetry breaking parameter:
\begin{equation}
\epsilon = \langle\phi_{WS}\rangle/M
\end{equation}
The extreme smallness of this parameter $\epsilon$ constitutes, of course, the
gauge hierarchy problem.

Here we are interested in the further
suppression of the quark and lepton mass
matrix elements relative to $\langle\phi_{WS}\rangle$.
We take the view \cite{fn1} that this
hierarchy strongly suggests the existence
of further approximately conserved chiral quantum numbers
beyond those of the SMG broken, say, by terms
of order ${\cal O}(\lambda)$. The SMG is then a
low energy remnant of some larger group $G$
and the fermion mass and mixing
hierarchies are consequences of the spontaneous
breaking of $G$ to the SMG.  The mass matrix element
suppression factors depend on how the fermions behave {\em w.r.t.\/} $G$
and on the symmetry breaking mechanism itself.

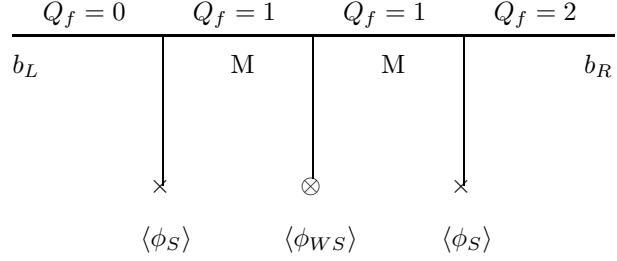
\begin{figure}[hbt]
\begin{center}
\setlength{\unitlength}{1mm}
\begin{picture}(80,40)(0,15)
\put(0,45){\line(1,0){80}}
\put(20,45){\line(0,-1){20}}
\put(40,45){\line(0,-1){19}}
\put(60,45){\line(0,-1){20}}
\put(18.3,24){$\times$}
\put(38.3,24){$\otimes$}
\put(58.3,24){$\times$}
\put(17,17){$\langle\phi_{S}\rangle$}
\put(36,17){$\langle\phi_{WS}\rangle$}
\put(57,17){$\langle\phi_{S}\rangle$}
\put(4,47){$Q_f = 0$}
\put(24,47){$Q_f = 1$}
\put(44,47){$Q_f = 1$}
\put(64,47){$Q_f = 2$}
\put(0,40){$\large{b}_L$}
\put(76,40){$\large{b}_R$}
\put(29,40){M}
\put(49,40){M}
\end{picture}
\end{center}
\par
\caption{Feynman diagram which generates the b quark mass via
superheavy intermediate states.}
\label{fig:pic}
\end{figure}

Consider, for example, an $SMG \times U(1)_f$ model,
whose fundamental mass scale is M, broken to the SMG
by the VEV of a scalar field $\phi_S$ where
$\langle\phi_S\rangle <  M$ and $\phi_S$
carries  $U(1)_f$ charge $Q_f(\phi_S)$ = 1.
Suppose further that $Q_f(\phi_{WS})=0$, $Q_f(b_L)=0$ and $Q_f(b_R)=2$.  Then
it is natural to expect the generation of a $b$ mass of order:
\begin{equation}
\left( \frac{\langle\phi_S\rangle }{M} \right)^2\langle\phi_{WS}\rangle
\end{equation}
via (see fig.~\ref{fig:pic}) the exchange of
two $\langle\phi_S\rangle$ tadpoles,
in addition to the usual
$\langle\phi_{WS}\rangle$ tadpole,
through two appropriately charged vector-like
superheavy (i.e. of mass M) fermion intermediate states \cite{fn1}.
Alternatively the transition can be mediated by a superheavy
Higgs state having $Q_f$ = 2.
We identify
$\epsilon_f=\langle\phi_S\rangle/M$
as the $U(1)_f$ flavour symmetry breaking parameter.
 In general we expect mass matrix elements of the form:
\begin{equation}
M(i,j) = \gamma_{ij} \epsilon_{f}^{n_{ij}}\langle\phi_{WS}\rangle,
\qquad \gamma_{ij} = {\cal O} (1),
\qquad n_{ij}= \mid Q_f(\psi_{L_{i}}) - Q_f(\psi_{R_{j}})\mid
\label{eq:mij}
\end{equation}
where $n_{ij}$
is the degree of forbiddenness due to the $U(1)_f$ quantum number difference
between the left- and right-handed fermion components. So the {\em effective\/}
SM Yukawa couplings of the quarks and leptons to the
Weinberg-Salam Higgs field
\begin{equation}
y_{ij} = \gamma_{ij}\epsilon_{f}^{n_{ij}}
\label{eq:yij}
\end{equation}
can consequently be small even though all
{\em fundamental\/} Yukawa couplings of
the ``true'' underlying theory are of $\cal O$(1).
We are implicitly  assuming here that
there exists a superheavy spectrum of states which can mediate all of the
symmetry breaking transitions; in particular we
do not postulate the {\em absence\/} of appropriate
superheavy states in order to obtain exact texture
zeros in the mass matrices \cite{dim}.
We now consider models based on this idea.

Recently a systematic analysis of fermion masses in SO(10) SUSY-GUT
models has been made \cite{anderson,raby} in terms of effective
operators obtained by integrating out the superheavy states, which
are presumed to belong to vector-like SO(10) {\bf16} + {$\bf \overline{16}$}
representations, in tree diagrams like fig.~\ref{fig:pic}. The minimal
number of effective operators contributing to mass matrices consistent
with the low energy data is four, which leads to the consideration of
GUT scale Yukawa coupling matrices satisfying Yukawa
unification, eq.~(\ref{yukun}), and having the following texture
\begin{equation}
Y_i = \pmatrix{0               & z_{i}'C   		 & 0\cr
		      z_{i}C  		  & y_{i}Ee^{i\phi}  & x_{i}'B\cr
		      0               & x_{i}B           & A\cr}
\end{equation}
where i = u, d, l. Here the $x_{i}$, $x_{i}'$, $y_{i}$, $z_{i}$
and $z_{i}'$ are SO(10) Clebsch Gordon coefficients. These Clebschs
can take on a very large number of discrete values, which are
determined once the set of 4 effective operators (tree diagrams)
is specified. A scan of millions of operators leads to just 9
solutions consistent with experiment, having Yukawa coupling matrices
with a partial Georgi-Jarlskog structure of the form:

\begin{equation}
Y_u =\pmatrix{0               & \frac{-1}{27}C   & 0\cr
		      \frac{-1}{27}C  & 0                & x_{u}'B\cr
		      0               & x_{u}B           & A\cr}
\qquad
Y_d =\pmatrix{0				  & C				 & 0\cr
			  C				  & Ee^{i\phi}		 & x_{d}'B\cr
			  0				  & x_{d}B			 & A\cr}
\qquad
Y_l =\pmatrix{0				  & C				 & 0\cr
			  C				  & 3Ee^{i\phi}		 & x_{l}'B\cr
			  0				  & x_{l}B			 & A\cr}
\end{equation}
 For each of the 9 models
the Clebschs $x_{i}$ and $x_{i}'$ have fixed values and
the Yukawa matrices depend on 6 free parameters: A, B, C, E,
$\phi$ and $\tan\beta$. The parameter hierarchy
$A \gg B$, $E \gg C$ and the texture zeros are
interpreted as due to an approximately conserved global $U(1)_f$ symmetry
and the chosen superheavy fermion spectrum.  The global $U(1)_f$
charges are assigned in such a way that only the 4 selected
tree diagrams are allowed. In particular the
texture zeros reflect the assumed absence of
superheavy fermion states which could mediate the transition between the
corresponding Weyl states.
Each solution gives 8 predictions consistent
with the data.

We now turn to models in which the chiral flavour charges are part of the
extended gauge group. The values of the chiral charges are
then strongly constrained by the anomaly conditions for the gauge theory.
It will also be assumed that any superheavy state needed to mediate a symmetry
breaking transition exists, so that the results
are insensitive to the details of the superheavy spectrum.
Consequently there will be no exact texture zeros but just highly suppressed
elements given by expressions like eq.~(\ref{eq:mij}). The aim in these
models is to reproduce all quark-lepton masses and mixing angles within
a factor of 2 or 3.

The $SMG \times U(1)_f$ model obtained by extending the SM with a gauged
abelian flavour group appears \cite{bijnens} unable to explain the
fermion masses and mixings using an anomaly-free set of flavour charges.
Models extending the SM (or the MSSM) with discrete gauge symmetries
and having new interactions at energies as low as 1 TeV have also been
investigated \cite{leurer}; as have gauged finite non-abelian
groups \cite{frampton}.

During the last year there has been considerable interest in models
\cite{ibanezross,binetruyramond,pokorski,nir,robinson}
extending the MSSM by a gauged abelian flavour group $U(1)_f$. The
abelian symmetry $U(1)_f$ is usually (see however \cite{jain})
assumed to be spontaneously broken by
\begin{enumerate}
\item[(a)] one chiral SMG-singlet field $\phi_S$ having a $U(1)_f$ charge
$Q_f(\phi_S) = -1$, or
\item[(b)] a vectorlike pair of SMG-singlet fields
$(\phi_S, \overline{\phi}_S)$ with charges $Q_f = (-1, +1)$,
which are assumed to acquire equal VEVs (along a D-flat direction).
\end{enumerate}
The effective MSSM Yukawa couplings $y_{ij}^{u,d,l}$ break the
$U(1)_f$ symmetry by
\begin{equation}
n_{ij}^{u,d,l} = Q_f(\psi_{L_{i}}) - Q_f(\psi_{R_{j}}) + Q_f(H_u, H_d, H_d)
\end{equation}
units of $U(1)_f$ charge, where $H_u$ and $H_d$ are the two MSSM
Higgs doublet fields. In case (a) the Yukawa coupling $y_{ij}^{u,d,l}$
will vanish whenever the excess charge $n_{ij}^{u,d,l}$ turns
out to be negative, due to the holomorphy of the the
superpotential which forbids the presence of the
conjugate field $\phi_S^{\dagger}$. For positive excess
charge the value of $n_{ij} > 0$ (omitting
the superscript $u,d,l$ for convenience) gives the degree of
suppression of the Yukawa coupling
$y_{ij} \sim \epsilon_f^{n_{ij}}$ as in
eq.~(\ref{eq:yij}). However it should be noted that the
supersymmetric zeros (for $n_{ij} < 0$) may be filled in,
by the wave function renormalisation of the kinetic
terms described by the K\"{a}hler potential, up to at most
a term of order $\epsilon_f^{|n_{ij}|}$. In case (b) the
low energy Yukawa couplings will be of order
$\epsilon_f^{|n_{ij}|}$ irrespective of the sign of
the excess charge.
Note also that if $Q_f(H) =  Q_f(H_u) + Q_f(H_d) \neq 0$, the
$\mu H_uH_d$ term is forbidden in the superpotential
\cite{nir,jain}. Furthermore if $Q_f(H)$ is moderately
negative, the $\mu$-term can arise from the K\"{a}hler potential
and naturally be of order the electroweak scale.

The new $U(1)_f$ gauge group is of course potentially
anomalous and the conditions for the cancellation of
anomalies provide strong constraints on the models.
Due to the possible presence of other SMG-singlet fields,
with non-zero $U(1)_f$ charges and no VEVs, it is assumed that
the $U(1)_f^3$ gauge anomaly and the mixed $U(1)_f$
gravitational anomaly are cancelled against such spectator
particles. However the conditions that the mixed $SU(3)^2U(1)_f$,
$SU(2)^2U(1)_f$, $U(1)_Y^2U(1)_f$ and $U(1)_f^2U(1)_Y$
gauge anomalies should vanish have only been satisfied in
one model \cite{pokorski} consistent with the phenomenological
mass ratios eq.~(\ref{eq:massratios}). This is a model of type (b),
with a vectorlike pair $(\phi_S, \overline{\phi}_S)$, having
the following $U(1)_f$ charge assignments (or an equivalent set)
for the quarks and leptons:
\begin{equation}
\left( \begin{array}{ccccc}
d_L \;&\; u_R \;&\; d_R \;&\; e_L \;&\; e_R \\
s_L \;&\; c_R \;&\; s_R \;&\; \mu_L \;&\; \mu_R \\
b_L \;&\; t_R \;&\; b_R \;&\; \tau_L \;&\; \tau_R
       \end{array} \right)
=\left( \begin{array}{ccccc}
 2 \;&\; -2  \;&\; 0 \;&\;  -5 \;&\;  -7 \\
 1 \;&\;  1 \;&\;  1 \;&\;  -3 \;&\;   5 \\
-1 \;&\;  3 \;&\;  1 \;&\;  -6  \;&\; -4
       \end{array} \right)
\label{eq:pokorski}
\end{equation}
The Higgs fields have $Q_f(H_u) = Q_f(H_d) =4$ and the corresponding
phenomenologically viable Yukawa matrices, generated using
eq.~(\ref{eq:yij}) with $\gamma_{ij} \simeq 1$ and
$\epsilon_f = \lambda$, have the form:
\begin{equation}
Y_u \simeq \pmatrix{\lambda^8  	   & \lambda^5    	& \lambda^3\cr
		    \lambda^9	   & \lambda^4   	& \lambda^2\cr
		    \lambda^{11}   & \lambda^2   	& 1\cr}
\quad
Y_d \simeq \lambda^2
	   \pmatrix{\lambda^4 	   & \lambda^3		& \lambda^3\cr
		    \lambda^5 	   & \lambda^2 		& \lambda^2\cr
		    \lambda^7 	   & 1			& 1\cr}
\quad
Y_l \simeq \lambda^2
	   \pmatrix{\lambda^4 	   & \lambda^3	 	& \lambda^2\cr
		    \lambda^5 	   & \lambda^2	  	& \lambda^3\cr
		    \lambda^2  	   & \lambda^5   	& 1\cr}
\label{eq:pok}
\end{equation}
For this model the ratio of the two MSSM Higgs VEVs is given by
$\tan\beta \simeq \lambda^2m_t/m_b \simeq 2$.

It is also possible to generate phenomenologically viable
Yukawa matrices with a single chiral $\phi_S$ field and
appropriate $U(1)_f$ assignments. However in such type (a)
models, the $SU(3)^2U(1)_f$, $SU(2)^2U(1)_f$ and
$U(1)_Y^2U(1)_f$ mixed anomalies no longer vanish
\cite{binetruyramond}; they can
only be cancelled by the Green-Schwarz mechanism
\cite{greenschwarz}, coming from an underlying string based
model. String theories in 4 dimensions contain an axion-like
field (the dilaton), which couples in a universal way to the
divergence of the anomalous currents. By assigning a $U(1)_f$
gauge variation to the dilaton it is possible to cancel the
anomalies, provided the mixed anomaly coefficients are in the
same ratio as the SU(3), SU(2) and $U(1)_Y$ Kac-Moody levels
$k_i$ in the string theory.
Gauge coupling unification at
the string scale requires the Kac-Moody levels and the
corresponding gauge coupling constants $g_i$ to satisfy
$k_3 g_3^2 = k_2 g_2^2 = k_1 g_1^2 = g_{\rm{string}}^2$.
The canonical normalisation of unified gauge coupling
constants $g_3^2 = g_2^2 = \frac{5}{3} g_1^2$ (corresponding to
the value $\sin^2 \theta_W = 3/8$ for the Weinberg angle)
at the string scale requires the Kac-Moody levels
to be in the ratio $k_3 : k_2 : k_1 = 1 : 1 : 5/3$.
In order that the Green Schwartz mechanism can function
properly, the mixed anomaly coefficients have to be in
the same ratio and it follows that \cite{binetruyramond,nir}
$\det Y_d/\det Y_l \simeq \lambda^{Q_f(H)}$. The fermion mass
ratios, eq.~(\ref{eq:massratios}), then require
$Q_f(H) = 0$ or $-1$. The Green-Schwartz mechanism
also requires \cite{dine} that the
$U(1)_f$ symmetry is spontaneously broken an order of
magnitude or so below the string scale.


In ref.~\cite{ibanezross}, Ibanez and Ross
considered type (b) models having symmetric mass matrices,
with a view to constructing an anomaly free
\mbox{$MSSM \times U(1)_f$} model having the texture
corresponding to one of the solutions in table~\ref{table:rrr}.
Different mass scales M and $\bar{M}$ are used in the two MSSM
Higgs sectors, giving two symmetry breaking parameters
$\epsilon = \langle\phi_{S}\rangle/M$ and
$\bar{\epsilon} = \langle\phi_{S}\rangle/\bar{M}$.
They took the two MSSM Higgs doublets to be neutral
under $U(1)_f$ and found a model similar to that of solution 2
as given in eq.~(\ref{eq:solution2}).
Again cancellation of the mixed anomalies of the $U(1)_f$ with
the SU(3), SU(2) and $U(1)_Y$ gauge groups is only possible
in the context of superstring theories via the
Green-Schwartz mechanism with $\sin^2 \theta_W = 3/8$.
The quarks and leptons are assigned the following $U(1)_f$ charges:
\begin{equation}
\left( \begin{array}{ccccc}
d_L \;&\; u_R \;&\; d_R \;&\; e_L \;&\; e_R \\
s_L \;&\; c_R \;&\; s_R \;&\; \mu_L \;&\; \mu_R \\
b_L \;&\; t_R \;&\; b_R \;&\; \tau_L \;&\; \tau_R
       \end{array} \right)
=\left( \begin{array}{ccccc}
-4 \;&\;  4 \;&\;  4 \;&\; -7/2 \;&\;  7/2 \\
 1 \;&\; -1 \;&\; -1 \;&\;  1/2 \;&\; -1/2 \\
 0 \;&\;  0 \;&\;  0 \;&\;   0  \;&\;  0
       \end{array} \right)
\label{eq:ross}
\end{equation}
which generate Yukawa matrices of the following form \cite{ibanezross}:
\begin{equation}
Y_u \simeq \pmatrix{\epsilon^8  	& \epsilon^3    & \epsilon^4\cr
		    \epsilon^3  	& \epsilon^2   	& \epsilon\cr
		    \epsilon^4  	& \epsilon   	& 1\cr}
\quad
Y_d \simeq \pmatrix{\bar{\epsilon}^8 & \bar{\epsilon}^3	& \bar{\epsilon}^4\cr
		    \bar{\epsilon}^3 & \bar{\epsilon}^2 & \bar{\epsilon}\cr
		    \bar{\epsilon}^4 & \bar{\epsilon}	& 1\cr}
\quad
Y_l \simeq \pmatrix{\bar{\epsilon}^7 & \bar{\epsilon}^3 & 0\cr
		    \bar{\epsilon}^3 & \bar{\epsilon}  	& 0\cr
		     0  	     & 0   		& 1\cr}
\end{equation}
There is a $Z_2$ symmetry forcing the (1,3), (3,1), (2,3) and
(3,2) elements of $Y_l$ to vanish. Alternatively the lepton charges
can be taken the same as the down quark sector giving $Y_l \simeq Y_d$.
The correct order of magnitude for all the masses and mixing angles are
obtained by fitting $\epsilon$, $\bar{\epsilon}$ and $\tan\beta$. This
is a large $\tan\beta \approx m_t/m_b$ model, but not necessarily
having exact Yukawa unification.

The fermion mass and mixing angle predictions from these models
could of course be more precisely tested if the $\cal O$(1)
coefficients $\gamma_{ij}$ in eq. (\ref{eq:yij}) were known.
It has recently been suggested \cite{lanzaross}
that they could be determined as infrared fixed point values
of the RGEs for the $MSSM \times U(1)_f$ model, supplemented by
a large number of
additional states in vector representations of the gauge group
which acquire their mass one or two orders of magnitude below
the string scale. In the presence of these extra states,
the gauge couplings are no longer asymptotically
free and the fundamental Yukawa couplings, being of $\cal O$(1),
evolve rapidly to their fixed point values just below the string scale.
Furthermore these vectorlike states are
assumed to form complete SU(5) representations (even though
the gauge group is not SU(5)), causing no relative evolution
of the SMG gauge couplings to leading order and increasing the
gauge unification scale close to the string scale. Encouraging
results have been obtained.

As discussed by Dudas and Savoy at this meeting, the role of
the $U(1)_f$ symmetry can be played by modular symmetry in
effective superstring theory \cite{dudas}. In this approach
the small parameter $\epsilon$ is identified as the
ratio of the real parts $Re T_\alpha$ of two moduli fields
and the $U(1)_f$ charges are replaced by the modular weights
of the fermion fields. This raises the possibility that the
whole structure of the fermion mass matrices could be determined
dynamically, by minimisation of the low energy effective potential
with respect to the moduli fields. There have also been some
recent developments on fermion masses in the
antigrand unification model \cite{fln:np1}
as discussed in Nielsen's talk.

\section{Conclusion}\label{sec:con}
We described two promising candidate dynamical calculations of the
top quark mass, using (a) the infrared quasifixed point value of
the MSSM renormalisation group equations, and (b) the multiple
point criticality principle in the pure Standard Model. In the
large $\tan\beta$ scenario all the third generation masses are
consistent with quasifixed point values and/or Yukawa unification.
However the mystery of the large top to bottom quark mass ratio
is then replaced by the mystery of a hierarchy of Higgs field
VEVs; also the possibly large SUSY radiative corrections for large
$\tan\beta$ make the mass predictions unreliable. There exist
several symmetric mass matrix ans\"{a}tze with texture zeros,
giving typically 5 successful relations between mass and mixing
parameters. Their hierarchical struture is most naturally
explained in terms of an approximate chiral (gauge) symmetry
beyond that of the Standard Model group. It is possible to
generate phenomenologically viable fermion mass matrices
in this way for the Minimal Supersymmetric Standard Model extended
by a spontaneously broken abelian flavour symmetry group.
In most cases this requires an underlying superstring model,
in which the Green-Schwartz mechanism
cancels anomalies with $\sin^2\theta_W =3/8$ at the string scale.



\end{document}